\documentclass[submitting]{nst}

\usepackage{subfigure,dcolumn}
\usepackage[T2A,T1]{fontenc}
\usepackage[russian,english]{babel}

\usepackage{listings}
\begin{document}

\title{Design of hadronic calorimeter for DarkSHINE experiment}\thanks{This work was supported by National Key R\&D Program of China (Grant No.: 2023YFA1606904 and 2023YFA1606900), National Natural Science Foundation of China (Grant No.: 12150006), and Shanghai Pilot Program for Basic Research—Shanghai Jiao Tong University (Grant No.: 21TQ1400209). }

\author{Zhen Wang}\thanks{Theses authors contributed equally to this work.}
\affiliation{Tsung-Dao Lee Institute, Shanghai Jiao Tong University, 1 Lisuo Road, Shanghai 201210, China}
\affiliation{Institute of Nuclear and Particle Physics, School of Physics and Astronomy, 800 Dongchuan Road, Shanghai 200240, China}
\affiliation{Key Laboratory for Particle Astrophysics and Cosmology (MOE), Shanghai Key Laboratory for Particle Physics and Cosmology (SKLPPC), Shanghai Jiao Tong University, 800 Dongchuan Road, Shanghai 200240, China}

\author{Rui Yuan}\thanks{Theses authors contributed equally to this work.}
\affiliation{Tsung-Dao Lee Institute, Shanghai Jiao Tong University, 1 Lisuo Road, Shanghai 201210, China}
\affiliation{Institute of Nuclear and Particle Physics, School of Physics and Astronomy, 800 Dongchuan Road, Shanghai 200240, China}
\affiliation{Key Laboratory for Particle Astrophysics and Cosmology (MOE), Shanghai Key Laboratory for Particle Physics and Cosmology (SKLPPC), Shanghai Jiao Tong University, 800 Dongchuan Road, Shanghai 200240, China}

\author{Han-Qing Liu}
\affiliation{Institute of Nuclear and Particle Physics, School of Physics and Astronomy, 800 Dongchuan Road, Shanghai 200240, China}
\affiliation{Key Laboratory for Particle Astrophysics and Cosmology (MOE), Shanghai Key Laboratory for Particle Physics and Cosmology (SKLPPC), Shanghai Jiao Tong University, 800 Dongchuan Road, Shanghai 200240, China}

\author{Jing Chen}
\affiliation{Institute of Nuclear and Particle Physics, School of Physics and Astronomy, 800 Dongchuan Road, Shanghai 200240, China}
\affiliation{Key Laboratory for Particle Astrophysics and Cosmology (MOE), Shanghai Key Laboratory for Particle Physics and Cosmology (SKLPPC), Shanghai Jiao Tong University, 800 Dongchuan Road, Shanghai 200240, China}
\affiliation{Tsung-Dao Lee Institute, Shanghai Jiao Tong University, 1 Lisuo Road, Shanghai 201210, China}

\author{Xiang Chen}
\affiliation{Institute of Nuclear and Particle Physics, School of Physics and Astronomy, 800 Dongchuan Road, Shanghai 200240, China}
\affiliation{Key Laboratory for Particle Astrophysics and Cosmology (MOE), Shanghai Key Laboratory for Particle Physics and Cosmology (SKLPPC), Shanghai Jiao Tong University, 800 Dongchuan Road, Shanghai 200240, China}

\author{Kim Siang Khaw}
\affiliation{Tsung-Dao Lee Institute, Shanghai Jiao Tong University, 1 Lisuo Road, Shanghai 201210, China}
\affiliation{Institute of Nuclear and Particle Physics, School of Physics and Astronomy, 800 Dongchuan Road, Shanghai 200240, China}
\affiliation{Key Laboratory for Particle Astrophysics and Cosmology (MOE), Shanghai Key Laboratory for Particle Physics and Cosmology (SKLPPC), Shanghai Jiao Tong University, 800 Dongchuan Road, Shanghai 200240, China}

\author{Liang Li}
\affiliation{Institute of Nuclear and Particle Physics, School of Physics and Astronomy, 800 Dongchuan Road, Shanghai 200240, China}
\affiliation{Key Laboratory for Particle Astrophysics and Cosmology (MOE), Shanghai Key Laboratory for Particle Physics and Cosmology (SKLPPC), Shanghai Jiao Tong University, 800 Dongchuan Road, Shanghai 200240, China}

\author{Shu Li}
\email[Corresponding author, ]{Shu Li, shuli@sjtu.edu.cn}{}
\affiliation{Tsung-Dao Lee Institute, Shanghai Jiao Tong University, 1 Lisuo Road, Shanghai 201210, China}
\affiliation{Institute of Nuclear and Particle Physics, School of Physics and Astronomy, 800 Dongchuan Road, Shanghai 200240, China}
\affiliation{Key Laboratory for Particle Astrophysics and Cosmology (MOE), Shanghai Key Laboratory for Particle Physics and Cosmology (SKLPPC), Shanghai Jiao Tong University, 800 Dongchuan Road, Shanghai 200240, China}

\author{Kun Liu}
\email[Corresponding author, ]{Kun Liu, kun.liu@sjtu.edu.cn}{}
\affiliation{Tsung-Dao Lee Institute, Shanghai Jiao Tong University, 1 Lisuo Road, Shanghai 201210, China}
\affiliation{Institute of Nuclear and Particle Physics, School of Physics and Astronomy, 800 Dongchuan Road, Shanghai 200240, China}
\affiliation{Key Laboratory for Particle Astrophysics and Cosmology (MOE), Shanghai Key Laboratory for Particle Physics and Cosmology (SKLPPC), Shanghai Jiao Tong University, 800 Dongchuan Road, Shanghai 200240, China}

\author{Qi-Bin Liu}
\affiliation{Tsung-Dao Lee Institute, Shanghai Jiao Tong University, 1 Lisuo Road, Shanghai 201210, China}
\affiliation{Institute of Nuclear and Particle Physics, School of Physics and Astronomy, 800 Dongchuan Road, Shanghai 200240, China}
\affiliation{Key Laboratory for Particle Astrophysics and Cosmology (MOE), Shanghai Key Laboratory for Particle Physics and Cosmology (SKLPPC), Shanghai Jiao Tong University, 800 Dongchuan Road, Shanghai 200240, China}

\author{Si-Yuan Song}
\affiliation{Institute of Nuclear and Particle Physics, School of Physics and Astronomy, 800 Dongchuan Road, Shanghai 200240, China}
\affiliation{Key Laboratory for Particle Astrophysics and Cosmology (MOE), Shanghai Key Laboratory for Particle Physics and Cosmology (SKLPPC), Shanghai Jiao Tong University, 800 Dongchuan Road, Shanghai 200240, China}
\affiliation{Tsung-Dao Lee Institute, Shanghai Jiao Tong University, 1 Lisuo Road, Shanghai 201210, China}

\author{Tong Sun}
\affiliation{Tsung-Dao Lee Institute, Shanghai Jiao Tong University, 1 Lisuo Road, Shanghai 201210, China}
\affiliation{Institute of Nuclear and Particle Physics, School of Physics and Astronomy, 800 Dongchuan Road, Shanghai 200240, China}
\affiliation{Key Laboratory for Particle Astrophysics and Cosmology (MOE), Shanghai Key Laboratory for Particle Physics and Cosmology (SKLPPC), Shanghai Jiao Tong University, 800 Dongchuan Road, Shanghai 200240, China}

\author{Xiao-Long Wang}
\affiliation{Key Laboratory of Nuclear Physics and Ion-beam Application (MOE), Fudan University, Shanghai 200443, China}
\affiliation{Institute of Modern Physics, Fudan University, Shanghai 200443, China}

\author{Yu-Feng Wang}
\affiliation{Tsung-Dao Lee Institute, Shanghai Jiao Tong University, 1 Lisuo Road, Shanghai 201210, China}
\affiliation{Institute of Nuclear and Particle Physics, School of Physics and Astronomy, 800 Dongchuan Road, Shanghai 200240, China}
\affiliation{Key Laboratory for Particle Astrophysics and Cosmology (MOE), Shanghai Key Laboratory for Particle Physics and Cosmology (SKLPPC), Shanghai Jiao Tong University, 800 Dongchuan Road, Shanghai 200240, China}

\author{Hai-Jun Yang}
\email[Corresponding author, ]{Haijun Yang, haijun.yang@sjtu.edu.cn}{}
\affiliation{Institute of Nuclear and Particle Physics, School of Physics and Astronomy, 800 Dongchuan Road, Shanghai 200240, China}
\affiliation{Key Laboratory for Particle Astrophysics and Cosmology (MOE), Shanghai Key Laboratory for Particle Physics and Cosmology (SKLPPC), Shanghai Jiao Tong University, 800 Dongchuan Road, Shanghai 200240, China}
\affiliation{Tsung-Dao Lee Institute, Shanghai Jiao Tong University, 1 Lisuo Road, Shanghai 201210, China}

\author{Jun-Hua Zhang}
\affiliation{Tsung-Dao Lee Institute, Shanghai Jiao Tong University, 1 Lisuo Road, Shanghai 201210, China}
\affiliation{Institute of Nuclear and Particle Physics, School of Physics and Astronomy, 800 Dongchuan Road, Shanghai 200240, China}
\affiliation{Key Laboratory for Particle Astrophysics and Cosmology (MOE), Shanghai Key Laboratory for Particle Physics and Cosmology (SKLPPC), Shanghai Jiao Tong University, 800 Dongchuan Road, Shanghai 200240, China}

\author{Yu-Lei Zhang}
\affiliation{Institute of Nuclear and Particle Physics, School of Physics and Astronomy, 800 Dongchuan Road, Shanghai 200240, China}
\affiliation{Key Laboratory for Particle Astrophysics and Cosmology (MOE), Shanghai Key Laboratory for Particle Physics and Cosmology (SKLPPC), Shanghai Jiao Tong University, 800 Dongchuan Road, Shanghai 200240, China}

\author{Zhi-Yu Zhao}
\affiliation{Tsung-Dao Lee Institute, Shanghai Jiao Tong University, 1 Lisuo Road, Shanghai 201210, China}
\affiliation{Institute of Nuclear and Particle Physics, School of Physics and Astronomy, 800 Dongchuan Road, Shanghai 200240, China}
\affiliation{Key Laboratory for Particle Astrophysics and Cosmology (MOE), Shanghai Key Laboratory for Particle Physics and Cosmology (SKLPPC), Shanghai Jiao Tong University, 800 Dongchuan Road, Shanghai 200240, China}

\author{Chun-Xiang Zhu}
\affiliation{Institute of Nuclear and Particle Physics, School of Physics and Astronomy, 800 Dongchuan Road, Shanghai 200240, China}
\affiliation{Key Laboratory for Particle Astrophysics and Cosmology (MOE), Shanghai Key Laboratory for Particle Physics and Cosmology (SKLPPC), Shanghai Jiao Tong University, 800 Dongchuan Road, Shanghai 200240, China}
\affiliation{Tsung-Dao Lee Institute, Shanghai Jiao Tong University, 1 Lisuo Road, Shanghai 201210, China}

\author{Xu-Liang Zhu}
\affiliation{Tsung-Dao Lee Institute, Shanghai Jiao Tong University, 1 Lisuo Road, Shanghai 201210, China}
\affiliation{Institute of Nuclear and Particle Physics, School of Physics and Astronomy, 800 Dongchuan Road, Shanghai 200240, China}
\affiliation{Key Laboratory for Particle Astrophysics and Cosmology (MOE), Shanghai Key Laboratory for Particle Physics and Cosmology (SKLPPC), Shanghai Jiao Tong University, 800 Dongchuan Road, Shanghai 200240, China}

\author{Yi-Fan Zhu}
\affiliation{Institute of Nuclear and Particle Physics, School of Physics and Astronomy, 800 Dongchuan Road, Shanghai 200240, China}
\affiliation{Key Laboratory for Particle Astrophysics and Cosmology (MOE), Shanghai Key Laboratory for Particle Physics and Cosmology (SKLPPC), Shanghai Jiao Tong University, 800 Dongchuan Road, Shanghai 200240, China}
\affiliation{Tsung-Dao Lee Institute, Shanghai Jiao Tong University, 1 Lisuo Road, Shanghai 201210, China}

\begin{abstract}
 The sensitivity of the dark photon search through final invisible decay states in low-background experiments relies significantly on the neutron and muon veto efficiencies, which depend on the amount of material used and the design of the detector geometry. This paper presents the optimized design of the hadronic calorimeter (HCAL) used in the DarkSHINE experiment, which is studied using a GEANT4-based simulation framework. The geometry is optimized by comparing a traditional design with uniform absorbers to one that uses different thicknesses at different locations on the detector, which enhances the efficiency of vetoing low-energy neutrons at the sub-GeV level. The overall size and total amount of material used in the HCAL are optimized to be lower, owing to the load and budget requirements, whereas the overall performance is studied to satisfy the physical objectives.
\end{abstract}


\maketitle
\flushbottom

\section{Introduction}
\label{sec:intro}

Over the past few decades, astronomical observations have indicated the presence of ordinary matter in the universe that can be observed by electromagnetic interactions, as well as a considerable amount of matter that does not interact with the electromagnetic force, which is called dark matter (DM)~\cite{Hooper:2007kb,Clowe:2006eq}. One can predict that dark matter not only interacts through gravitational force but can also be studied as dark matter particles from a particle physics perspective. Research conducted from this perspective offers a mechanism to elucidate the evolutionary process of dark matter composition and to investigate potential novel interactions between DM candidate particles and Standard Model (SM) particles. \\

Within a class of prevailing theories, a mechanism known as ``freeze-out''~\cite{Du:2021jcj} has been introduced to elucidate the evolutionary process. The universe was in thermal equilibrium at the beginning of its existence, and dark matter was constantly created and annihilated in pairs. As the universe subsequently underwent expansion and cooling, resulting in a greater dispersion of matter, the density of dark matter reached a steady state. ``Freeze-out'' facilitates the existence of dark matter across a broad mass spectrum ranging from MeV to 10s~TeV, and which can be further divided into light dark matter (LDM) within the MeV to GeV mass range and weakly interactive massive particles (WIMP) within the GeV to TeV mass range~\cite{Griest:1989wd,Ho:2012ug,Steigman:2013yua,Boehm:2013jpa,Nollett:2013pwa,Nollett:2014lwa,Serpico:2004nm}. \\

The search for dark matter particles is essential and challenging in elementary particle physics. Among these, WIMPs have been extensively explored for a long time. In general, the WIMP hypothesis provides a more natural and intuitive framework for the existence and detection of large mass but weakly interacting particles. Numerous experiments have obtained constraints on the mass of WIMPs~\cite{Liu:2017drf}, such as AMS~\cite{Giovacchini:2020vxz}, DAMPE~\cite{Kyratzis:2022gvg}, LHC~\cite{LHC}, BESIII~\cite{Prasad:2019ris}, XENON~\cite{XENON:2018voc,XENON:2023cxc}, and PandaX~\cite{PandaX-4T:2021bab}. These experiments encompass space, collision, and underground experiments, and search for WIMPs through direct and indirect exploration. Thus far, the mass limits of WIMPs have been close to the neutrino floor~\cite{Billard:2021uyg}. \\

However, current research on LDM remains insufficient, rendering it a prominent subject in recent investigations of dark matter. Beyond the Standard Model (SM) theory of the LDM, a particle analogous to an ordinary electromagnetic photon is introduced as a mediator for transporting the interaction between dark matter, commonly referred to as a dark photon ($A'$)~\cite{Holdom:1985ag,Foot:1991kb}. Moreover, a dark photon can be coupled to an SM photon via kinetic mixing ($\epsilon$) and subsequently interacts with SM particles. Dark photons play a crucial role in mediating interactions between dark and ordinary matter~\cite{Fuyuto:2019vfe,Choi:2020pyy,Cheng:2021qbl}. Several international experiments, including NA64~\cite{NA64:2023wbi}, BESIII~\cite{Zhang:2019wnz}, and LDMX~\cite{LDMX:2018cma,LDMX:2019gvz,Berlin:2018bsc}, are currently in operation or are under development to search for dark photons. \\

\begin{figure}[htb]
\centering
\includegraphics
  [width=0.9\hsize]
  {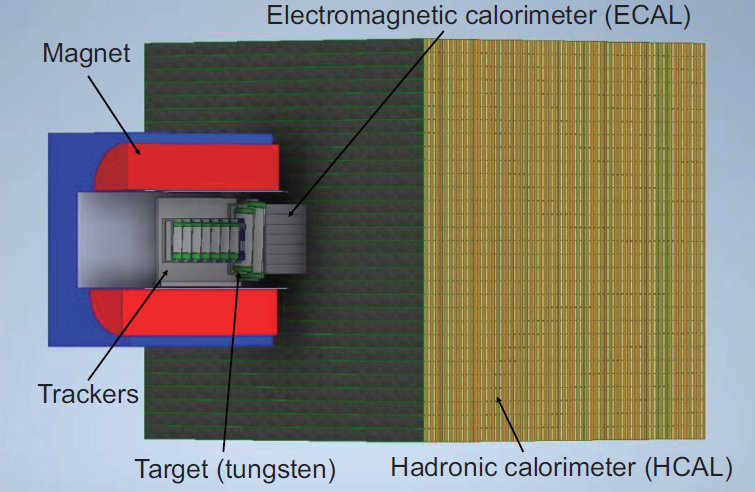}
\caption{Schematic of DarkSHINE detector. Electron
incident direction is from left to right, and the red material with a blue brace is the dipole magnet. The tagging tracker is placed at the center of it, while the recoil tracker is located at the edge of the magnet. The target is caught in the middle of the tracker. ECAL is placed after the tracker, followed by HCAL.~\cite{Chen:2022liu}}
\label{fig:darksine}
\end{figure}

The DarkSHINE experiment~\cite{Chen:2022liu,SLi} is a new initiative of fixed-target experiments that utilizes an 8 GeV high repetition rate low-current electron beam, which will be provided by the Shanghai High Repetition-Rate XFEL and Extreme Light Facility (SHINE)~\cite{Wan:2022het,Zhao:2017ood,Zhao:2016ood}. Their primary goal is to search for dark photons through their invisible decay into dark matter particles. DarkSHINE is expected to exhibit sensitivity to dark photons within the mass range of MeV–GeV. An illustrative overview of the preliminary design of the DarkSHINE detector is shown in Figure~\ref{fig:darksine}, including a silicon tracker, electromagnetic calorimeter (ECAL), and hadronic calorimeter (HCAL). A fixed tungsten target is placed between the tagging and recoil trackers, and immersed in a magnetic field. \\

The silicon tracker, immersed in a 1.5~T magnetic field generated by the magnet system, is used to reconstruct the trajectory of the incident and recoil electrons and obtain the momentum of the electrons. The tracker system includes a tagging tracker and recoil tracker, both of which are immersed in a magnetic field. The tagging tracker comprises seven silicon strip layers, whereas the recoil tracker has six layers. A tungsten~(W) target is placed between two parts of the tracker. The target has a decay length of 0.1$X_0$. In each layer of the tracking module, two silicon strip sensors are placed at a small angle (100mrad) to improve position accuracy. \\

ECAL is placed after the recoil tracker and comprises 11 layers of crystal scintillator. Each layer includes a 20$\times$20 LYSO~(Ce) crystal scintillator with an area of 2.5~cm $\times$2.5~cm and a length of 4cm. Eleven layers provide 44 $X_0$ of decay length. The design aims to effectively absorb all the energy of the incoming electrons and photons while utilizing crystals with excellent energy resolution to achieve optimal sensitivity. In addition to enabling more precise measurements of the deposition energy, the combination of information captured by ECAL and the tracker facilitates a comprehensive reconstruction of recoil electrons. \\

\begin{figure}[htb]
\centering
\includegraphics
  [width=0.9\hsize]
  {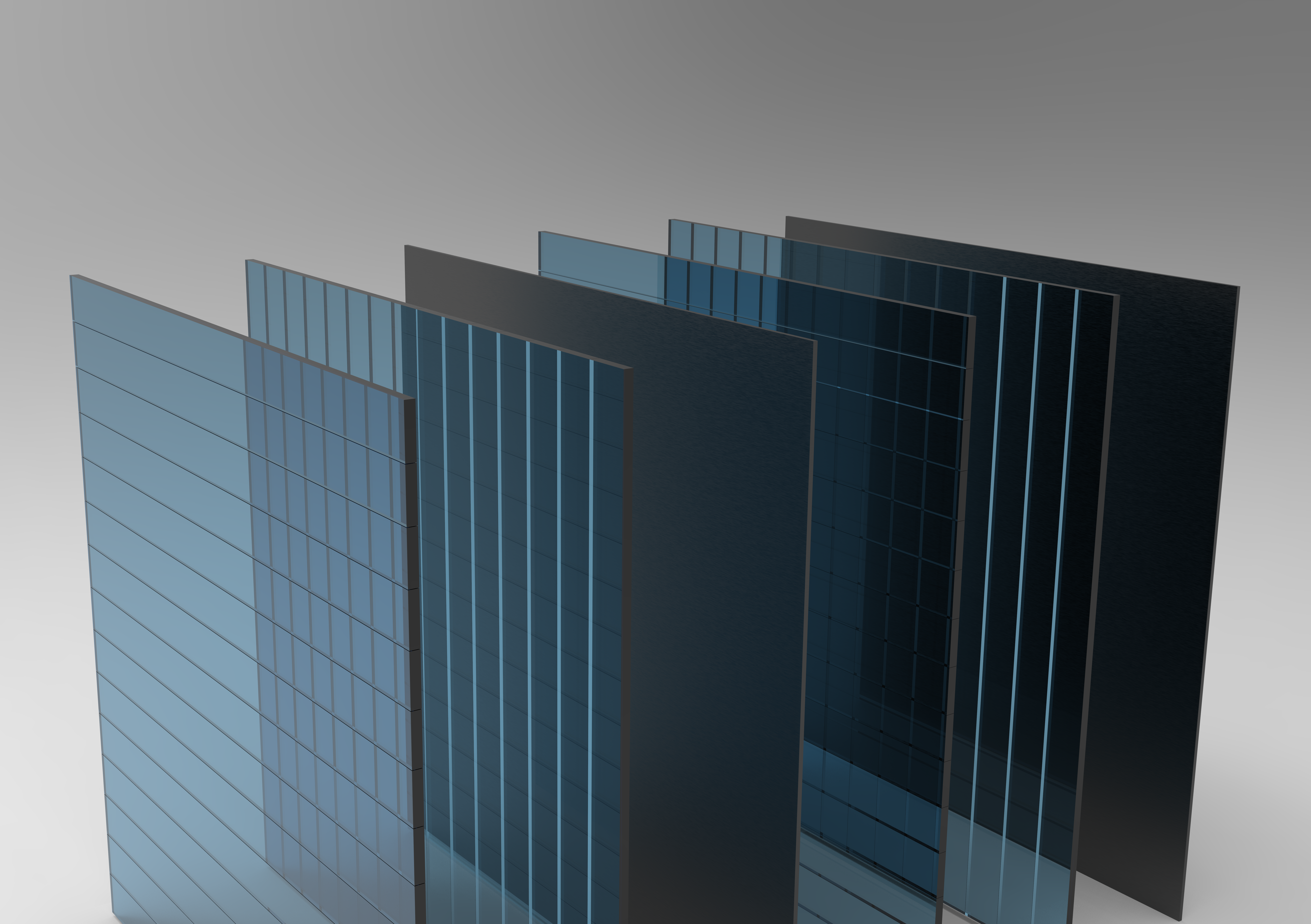}
\caption{Schematic of the xy-crossing scintillator structures. Transparent parts are scintillator layers and the opaque parts are iron layers. One scintillator layer comprises x-direction sub-layer and y-direction sub-layer.}
\label{fig:xy_crossing}
\end{figure}

HCAL is a sampling calorimeter of ``Fe-Sc'' type, which uses iron as the absorber layer and plastic scintillator strips to construct the sensitive layer. Each sensitive layer comprises two scintillator layers positioned between two absorbers. The scintillator strips in the two layers are perpendicular to each other in the $xy$-plane, as illustrated in Figure~\ref{fig:xy_crossing}. The cross-sectional area of HCAL is 4~m$\times$4~m, and the length is 4~m; thus, these materials can provide more than 10 $\lambda$ of interaction length.  \\

\begin{figure}[htb]
\centering
\includegraphics
  [width=0.9\hsize]
  {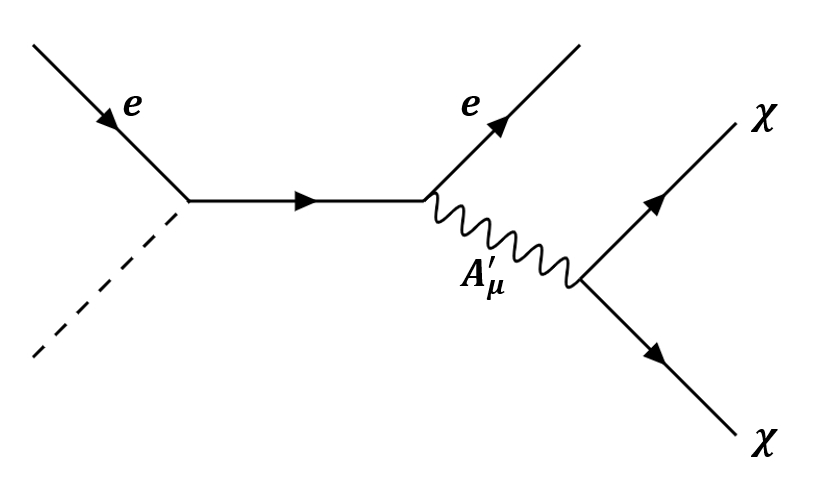}
\caption{Feynman diagram illustrating the signal process in the DarkSHINE experiment, encompassing bremsstrahlung production and invisible decay of dark photons~\cite{Chen:2022liu}.}
\label{fig:feynman_dia}
\end{figure}

The interaction between the electron and target is expected to produce dark photons, as illustrated in Figure~\ref{fig:feynman_dia}. These dark photons carry a portion of the incident electron energy and subsequently decay into dark matter particles, which traverse the remaining detectors without leaving discernible traces. Simultaneously, the residual energy is carried away by the recoil electron, traversing a path within the recoil tracker, and is subsequently fully absorbed by the ECAL. In certain scenarios, dark photons may exhibit a visible decay mode. The decay of dark photons into a pair of SM particles introduces an additional vertex containing $\epsilon$, resulting in a significantly suppressed cross-section for this process compared to that of invisible decay~\cite{Chen:2022liu}. In this invisible decay signal process, the energy difference between the incident and deposited energies of the ECAL can be treated as the energy of the dark photons. \\ 

HCAL is designed to veto background events that exhibit similar behavior in the tracker and ECAL as signal events. These events typically involve neutrons and muons, which can occur in both the target and ECAL areas, resulting in minimal energy deposition in ECAL but a detectable number of deposits in HCAL. Because ECAL functions as a fully absorbed calorimeter, neither the recoiled electrons nor the dark matter provide any discernible information in HCAL. Therefore, when HCAL registers a certain amount of deposited energy, it serves as a veto condition for this type of background scenario. As a low-background experiment, the sensitivity of DarkSHINE is directly influenced by the power of HCAL to identify these events. \\

This paper presents the design and optimization of HCAL for DarkSHINE experiments. The primary criterion for evaluating the optimization is the capability of the detector to discern events containing neutrons, which are predominant in target particles and pose challenges for detection. The need for optimization mainly arises from budget constraints and building load considerations. It is necessary to reduce the weight of HCAL within a specific range while ensuring that an adequate amount of material is used to minimize waste. The weight is dominated by iron, and the cost is mainly owing to the scintillator, which is directly related to the number of layers. Section ~\ref{sec:background_process} provides an overview of the background processes related to this optimization concern and discusses the optimization criteria. Sections ~\ref{sec:optimization} and~\ref{sec:sideHcal} present the details and results of the optimization, respectively. Finally, Section ~\ref{sec:conclusion} provides a comprehensive summary. \\

\section{Main Background in HCAL and treatment}
\label{sec:background_process}

Beam electrons mostly pass through the tungsten target without any interaction, predominantly depositing the bulk of their energy in the ECAL in the form of an electromagnetic shower. The rejection of these background events is straightforward; a cut-off based on the total energy accumulated in the ECAL~($\mathrm{E_{ECAL}}$) can be employed because recoil electrons from signal processes are expected to possess lower deposited energies~\cite{Chen:2022liu}. However, a small fraction of electrons generate an additional photon through the process of hard bremsstrahlung emissions. These bremsstrahlung photons can either end in electromagnetic showers within ECAL and be vetoed using a similar $\mathrm{E_{ECAL}}$ cut, or exhibit conversions into lepton or hadron pairs, which may occur within both the target and ECAL. In the context of electron pairs, events can be identified either in the tracker and/or ECAL, depending on the location of the conversions. \\

The HCAL plays a crucial role in the conversion of photons into $\mu$ pairs and hadron pairs~\cite{LDMX:2023zbn}. Muons pass through the ECAL as minimum ionizing particles~(MIP), which reduces the effectiveness of the $\mathrm{E_{ECAL}}$ cut. While DarkSHINE ECAL provides additional information, such as tracks and topology within the ECAL, it is essential to emphasize that the information obtained from HCAL remains paramount and straightforward without requiring complex reconstruction algorithms. This situation is analogous to the final states of charged hadron pairs, where combining information from both ECAL and HCAL can result in exclusion. However, in the case of neutral hadrons that do not decay within ECAL, discrimination power relies heavily on HCAL. \\

These bremsstrahlung photons can also interact with materials within the target and ECAL, resulting in photon-nuclear reactions that facilitate the rise of neutral hadrons. There exists a class of processes that is significantly less frequent~\cite{LDMX:2023zbn}, exhibiting behavior similar to that of the signal processes in the tracker and ECAL. In these cases, a single energetic neutral hadron is typically involved. The identification of these particles involves subjecting them to materials that induce hadronic showers, encompassing both QED and QCD~\cite{qcd1,qcd2,qcd3} processes. It can be predicted that an HCAL with sufficient absorber thickness can capture a certain amount of the shower energy and veto such events. Furthermore, electron-nuclear interactions between the materials of ECAL and the target also involve nucleon production, and the treatment remains the same as that for photon-nuclear processes. \\

\begin{figure}[htb]
\centering
\includegraphics
  [width=0.9\hsize]
  {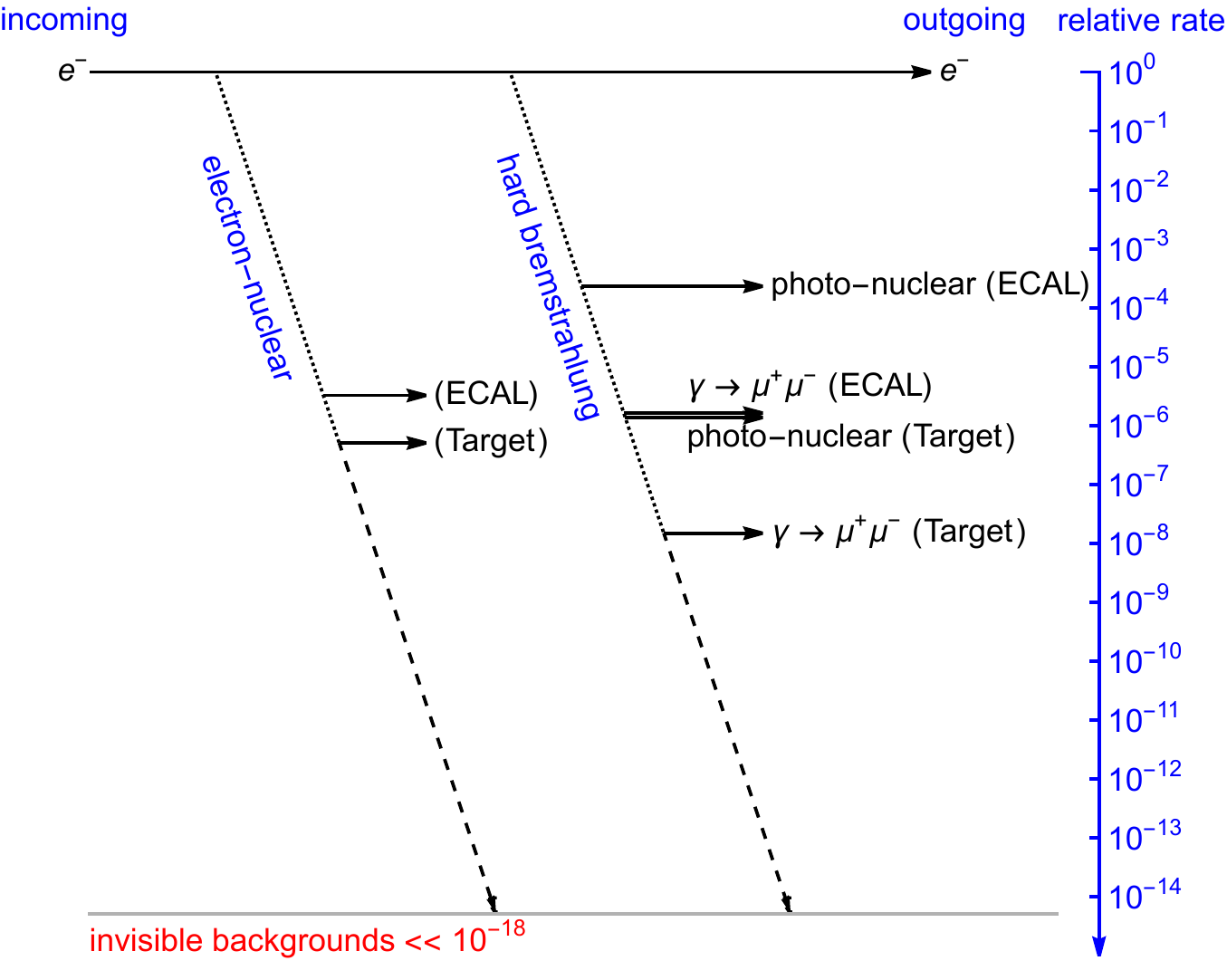}
\caption{Flow of background processes, ECAL and target refer to the locations where the processes occur.}
\label{fig:flowbackground}
\end{figure}

A schematic of these processes is presented in Figure~\ref{fig:flowbackground}. The previously discussed background can be suppressed by the information provided by the detector, and its detection efficiency imposes limitations on the sensitivity of the experiments. Further, there is an irreducible physics background that encompasses neutrino processes. Two leading irreducible reactions exist ~\cite{Chen:2022liu,Izaguirre:2014bca}. The first is Moller scattering, followed by the charged-current quasi-elastic (CCQE) reaction $e^{-}p \rightarrow \nu_{e}n$. The subsequent process involves neutrino pair production $e^{-}N \rightarrow e^{-}N\nu\bar{\nu}$. These neutrinos account for a large fraction of incident electron energy. However, these backgrounds exhibit a rate approximately four orders of magnitude~\cite{Chen:2022liu,LDMX:2023zbn} lower than the 3$\times$10$^{14}$ electron-on-target ~(EOT) and will not be addressed in this study owing to their negligible impact. \\

Compared with neutral hadrons, muons are easier to detect because of their significant energy deposits in sufficient layers of scintillators. The primary concern lies in the HCAL's capability of to detect neutral hadrons, as the veto power of hadronic particles is a crucial function and design consideration. These rare processes, such as photon-nuclear (PN) and electron-nuclear (EN) reactions, can be further categorized based on whether they occur in the target or ECAL. These are referred to as PN-target, PN-ECAL, EN target, and EN-ECAL, respectively. Table~\ref{tab:background_process} summarizes the particles most frequently generated from these rare processes. Given that neutrons constitute the largest proportion and protons can be excluded through a combination of tracker and ECAL information, this study primarily utilized neutrons to validate the optimization effect, with all target particles subsequently tested. \\

\begin{table}[!htb]
\caption{Particle types and frequencies from electron-nuclear and photon-nuclear process, neutrons are predominant.
\label{tab:background_process}}
\centering
\vspace{0.3cm}
\begin{tabular}{l|l|l|l|l}
\toprule
Process         & Neutron & Proton  & Pion    & Kaon   \\
\midrule
Electron-Nuclear & 73.42\% & 21.52\% & 4.64\%  & 0.42\% \\
Photon-Nuclear   & 64.95\% & 18.56\% & 14.43\% & 2.06\% \\
\bottomrule
\end{tabular}
\end{table}

HCAL rejects an event by setting cuts on the deposited energy of the neutrons in the event, and the efficiency of the veto varies for neutrons of different energies. Notably, the veto efficiency of a single-neutron event is identical to that of a neutron, whereas the efficiency of a multi-neutron event is equivalent to the veto efficiency of at least one of these neutrons. This veto performance is evaluated based on a number defined as the ratio between the number of events (or neutrons) that are not vetoed and the total number, which is referred to as veto inefficiency. \\

\begin{figure}[htb]
\centering
\includegraphics
  [width=0.9\hsize]
  {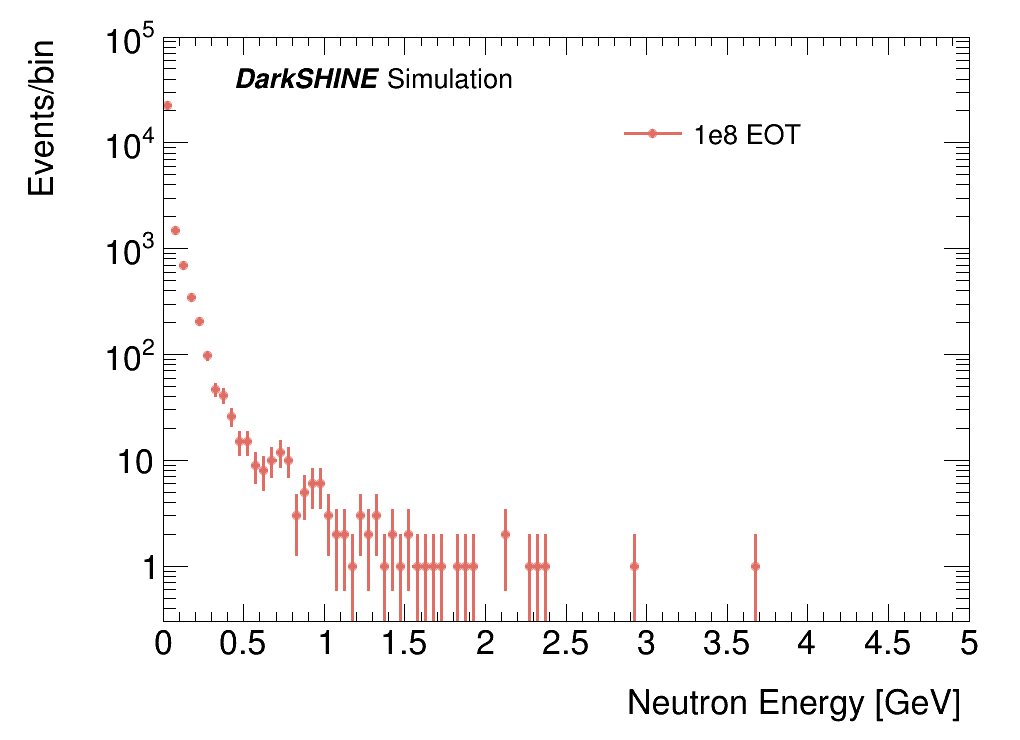}
\caption{Neutron energy distribution after applying cut on ECAL energy to request  $\mathrm{E_{ECAL}} < 2.5$~GeV. The result shows that very few neutrons with energy exceeding 2.5~GeV are left after ECAL cut. Thus, under assumptions of 1$\times$10$^{14}$ EOT, at most 1$\times$10$^{6}$ level energetic neutrons are expected.}
\label{fig:neutron_energy}
\end{figure}

The energy distribution and the number of neutrons in the predicted events are studied, as illustrated in Figure~\ref{fig:neutron_energy}. As discussed in Section~\ref{sec:intro}, the ECAL of DarkSHINE absorbs all photon and electron energy, providing the total deposited energy quantity. The variable $\mathrm{E_{ECAL}}$ can effectively discriminate against numerous background events, because the majority of background events tend to exhibit higher values of $\mathrm{E_{ECAL}}$ than the signal. To specifically focus on events that cannot be rejected by other subdetectors but rely on the rejection power of the HCAL, only events satisfying the cut $\mathrm{E_{ECAL}} < 2.5$~GeV~\cite{Chen:2022liu} are presented. \\

This study involves simulating 1$\times$10$^{8}$ electrons hitting the target. Considering that the ECAL energy loss is approximately 5.5 GeV, even if there is a neutron with an energy of approximately 2.5 GeV in the event, which accounts for approximately half of the energy loss, it can still be inferred that the event comprises other components with similar energies. Consequently, events involving a single high-energy neutron contribute significantly to the overall energy loss, whereas the remaining components lack sufficient energy for veto and are rare. To assess the quality of the optimization results, neutrons around 2GeV are considered as energetic particles. \\

Considering that only a few neutrons survive in this phase space, it is expected that approximately 1$\times$10$^{6}$ energetic neutrons will be generated under the conditions of a 1$\times$10$^{14}$ EOT, which is consistent with the predicted number to be collected within one year~\cite{Chen:2022liu}. Consequently, a veto inefficiency of < 10$^{-5}$ is selected as the performance benchmark for high-energy neutron rejection, which is capable of reducing energetic neutrons to the unit level. Conversely, in the absence of high-energy neutrons and the presence of only low-energy neutrons, it is implausible that these particles are the sole particles detected in the event; otherwise, the ECAL would have recorded an energy deposition closer to 8GeV. In such a scenario, a veto inefficiency of < 10$^{-3}$ would suffice to achieve an equivalent rejection power if multiple neutrons are present.

\section{Optimization of the design}
\label{sec:optimization}

\subsection{Simulation introduction}
\label{sec:simulation_framework}

The optimization study is conducted using DarkSHINE software~\cite{Chen:2022liu}, which is a comprehensive simulation and analysis framework that seamlessly integrates various functions such as detector simulation, electronic signal digitization, event display, event reconstruction, and data analysis. In an upcoming iteration, machine learning, which is becoming increasingly prevalent in high-energy physics research~\cite{He:2023zin}, will be implemented. This all-in-one package, built on GEANT4 v10.6.3~\cite{GEANT4:2002zbu}, is characterized by a DarkSHINE detector, and employs an internal data structure to facilitate efficient data flow across different stages. \\

GEANT4 is a comprehensive toolkit specifically designed to simulate the interaction of particles with matter, and plays a critical role in the development and optimization of HCAL. It is widely used in high-energy physics analyses owing to its extensive functionalities including particle tracking, detailed geometry configurations, sophisticated physics models, and particle hit detection. The toolkit supports a wide range of physical processes encompassing electromagnetic, hadronic, and optical interactions. Further, it provides an extensive library of long-lived particles, materials, and elements across various energy spectra. The architecture of GEANT4 excels in managing complex geometries while offering flexibility for customization to satisfy the unique demands of scientific and engineering applications. \\

The properties of GEANT4 facilitate the employment of two different simulation strategies. First, the target particles are simulated directly to hit the HCAL without considering other detector components. This approach allows us to obtain large statistics easily and intuitively calculate the veto efficiency of a target particle(Section~\ref{sec:optimization}). Second, complete simulations are conducted to evaluate the veto efficiency of rare process events involving neutrons or muons by applying bias functions while electrons interact with the target and traverse all the detectors (Section~\ref{sec:sideHcal}). The absence of this bias function results in excessive consumption of computing resources when simulating inclusive background instances until the statistics of certain rare process instances reach the required level. \\

As documented in ~\cite{Chen:2022liu}, the DarkSHINE experiment employs two cuts to veto the muons and hadronic particles, relying on the HCAL variables, $\mathrm{E_{HCAL}^{total}}$ and $\mathrm{E_{HCAL}^{MaxCell}}$. Here, $\mathrm{E_{HCAL}^{total}}$ represents the total energy collected in the HCAL and $\mathrm{E_{HCAL}^{MaxCell}}$ corresponds to the highest energy deposition among all the cells(scintillator strip). In this study, considering the wide range of sizes tested, a scintillator unit width of 5cm is selected because of its small and easily divisible value. \\

The cut value has been optimized since its initial publication~\cite{Chen:2022liu} to achieve an improved signal-to-background ratio. Therefore is it adopted as the baseline selection criterion in this study:
\begin{itemize}
    \item total energy reconstructed in HCAL, $\mathrm{E_{HCAL}^{total}}<30$~MeV;
    \item maximum cell energy in HCAL, $\mathrm{E_{HCAL}^{MaxCell}}<0.1$~MeV.
\end{itemize}

To achieve a sufficiently accurate estimation to satisfy the requirement of < 10$^{-5}$ veto inefficiency, the studies presented in this section employ 10$^6$ events for each test point(excluding section~\ref{sec:performance}, wherein the number is 10$^7$). In the corresponding plots, even if 0 out of 10$^6$ events survive the cuts, they are still counted as 1$\times$10$^{-6}$ veto inefficiencies, which is identical to the scenario where 1 out of 10$^6$ events survives. This treatment is necessary because the current number of simulated events cannot represent a scenario between 0 and 1$\times$10$^{-6}$ while maintaining the validity of the logarithmic axis. \\

\subsection{Transverse Size} 
\label{sec:Transverse}

The transverse dimensions of the HCAL, which determine its coverage angle, are crucial for effectively vetoing events with neutral particles and analyzing rare processes. The consideration of the coverage angle encompasses both the trajectories of secondary particles generated from electron-target interactions and the dimensions of both electromagnetic and hadronic showers. When designing and optimizing an HCAL, a comprehensive range of factors must be considered. Although a larger size offers benefits, its use is limited. Therefore, considering the physical requirements, budget costs, and weight limitations imposed by the experimental conditions become the dominant factor. As a sampling calorimeter, it is not essential to capture complete shower information. Instead, it only requires sufficient components within the shower cluster to be deposited in the scintillator and subsequently vetoed by the designated selection criteria. Therefore, provided that the exclusion efficiency of the HCAL satisfies the criteria discussed in Section~\ref{sec:background_process}, opting for a smaller size would yield a reduced budget and overall weight, thereby constituting an optimal choice. \\

Diverse detectors exist worldwide with varying coverage angles designed for specific purposes. Our methodology focuses on monitoring secondary particles projected forward in alignment with the direction of the beam. By subjecting the HCAL to direct tests involving the injection of various particles, its performance can be independently evaluated because it serves as the final component of the detector. Considering that the incident is a particle rather than a complete event, it is unnecessary to introduce the screening information provided by the other subdetectors at this stage. This approach facilitates the accumulation of experience in HCAL design, rather than focusing on a highly specific case. Moreover, it enables a comprehensive understanding of how the lateral size of the HCAL impacts efficiency and effectiveness across various experimental scenarios. \\

This study meticulously explores the influence of varying transverse sizes on the veto capabilities of an HCAL, spanning dimensions from 4~m$\times$4~m~\cite{Chen:2022liu} to a more compact 1~m$\times$1~m scale. Simulations were meticulously conducted for each specified transverse size, adhering to the methodological framework described in Section ~\ref{sec:simulation_framework}. To comprehensively investigate the variation trends, a range of particle energies from 100 MeV to 3000 MeV and several size options were traversed. A pivotal aspect of this study is the maintenance of a constant total absorber thickness, precisely $~10 \lambda$, across all designs. This strategic decision is aimed at mitigating any potential bias that might arise from variations in the total detector thickness, thereby ensuring that the observed differences in veto capabilities can be attributed solely to the transverse dimensions. \\

The HCAL design under investigation comprises four strategically arranged modules in a two-by-two configuration. These modules were meticulously dimensioned to be half the length of the HCAL's transverse side, reflecting a deliberate design choice that effectively balanced structural integrity and functional efficiency. Moreover, the scintillator strips, which are essential for the detection capabilities, were designed with a length equivalent to half of the HCAL's transverse side while maintaining a consistent width of 5 cm. This specification ensures enhanced sensitivity while keeping the detector size manageable. Furthermore, this modular design will facilitate future detector construction and installation processes, including the placement of readout electronics and design of support structures. \\

\begin{figure}[htb]
\centering
\includegraphics
  [width=0.9\hsize]
  {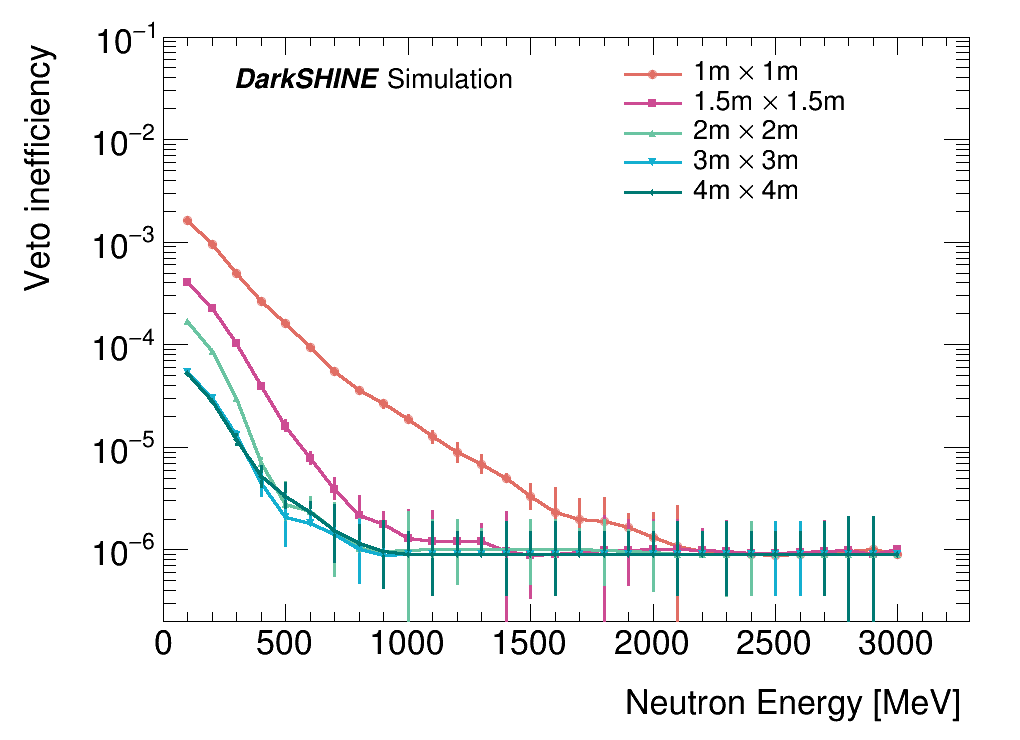}
\caption{Veto inefficiency as a function of different incident neutron energies. Larger size HCAL exhibits better veto power as expected owing to its capability of acceptance compared with smaller size HCAL designs. To satisfy the weighting limits of the SHINE facility, the 1.5~m design is selected to be the final choice.}
\label{fig:horizontal_size}
\end{figure}

The results of these simulations, illustrated in Figure~\ref{fig:horizontal_size}, provide critical insights into the relationship between the HCAL's transverse size and its veto efficiency. The Y-axis represents the veto inefficiency, each curve corresponds to a specific size choice, and the x-axis denotes the incident particle energy. Preliminary findings suggest that variations in transverse dimensions significantly affect the HCAL's ability to veto background events effectively. \\

In conclusion, these designs demonstrate equivalent veto power for high-energy neutrons in the energy range of 2–3~GeV, while the performance of 1~m$\times$1~m design deteriorates significantly between 1–2~GeV compared to the design above 1.5~m. For low-energy neutrons, larger area configurations offer enhanced performance. The veto inefficiency of low-energy neutrons in the 1.5~m$\times$1.5~m design was already less than 10$^{-3}$, which satisfied the specified requirement. Furthermore, the disparity in the low-energy neutron veto between the 1.5 and 1~m designs was more pronounced than that between the 4 and 1.5~m designs. Given the constraints of the SHINE facility on the supporting structure and weight, a 1.5~m$\times$1.5~m design was selected to ensure sufficient interaction length while minimizing weight. \\

\subsection{Absorber thickness}
\label{sec:Absorber}

Depending on the experimental design and specific location of the detector within the overall experiment, there can be significant variations in the required information, leading to diverse approaches for designing the detector, even when employing identical materials. Among them, homogeneous sampling detectors stand out because of their distinct design principles and applications. Homogeneous detectors are characterized by their uniform composition and utilize a single material that simultaneously acts as both the active medium for detecting particles and the absorber. This design ensures a high resolution for measuring the energy of incoming particles, making homogeneous detectors particularly useful in environments where precision is paramount. The sampling detectors are constructed from alternating layers of active and passive materials. The active layers are responsible for detecting particles, whereas the passive layers absorb them, thereby facilitating the measurement of particle energies. Although sampling detectors may offer lower resolution than their homogeneous counterparts, they are highly valued for their efficiency and versatility in handling high-energy particles and complex events. \\

The DarkSHINE HCAL detector functions as a sampling detector, capturing only a portion of the incoming energy in the sensitive layers. Consequently, an ideal absorber should be capable of capturing high-energy neutrons, while retaining sufficient energy for low-energy neutrons to reach the scintillators. In this study, the hadronic veto system effectively detected neutrons ranging from approximately 100 MeV to a few GeV. Low-energy neutrons would rapidly lose their energy within the absorbers, depositing minimal energy in the scintillator, whereas detectors may easily overlook high-energy neutrons if the absorber's thickness is inadequate. \\

\begin{figure*}[htb]
\centering
\includegraphics
  [width=0.45\hsize]
  {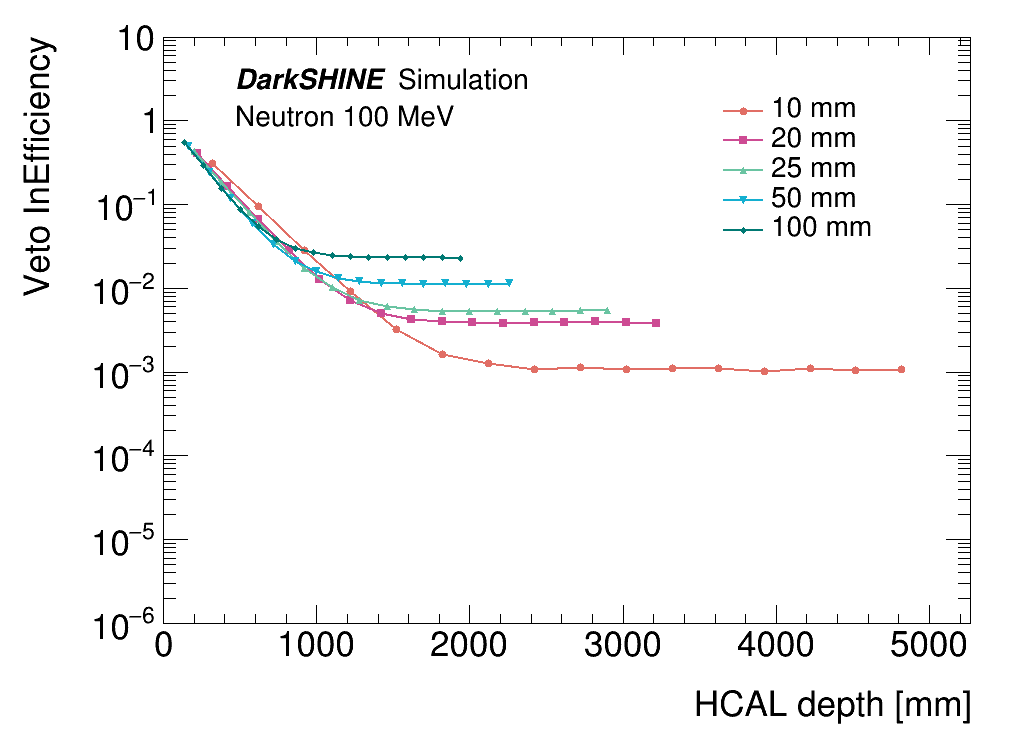}
\includegraphics
  [width=0.45\hsize]
  {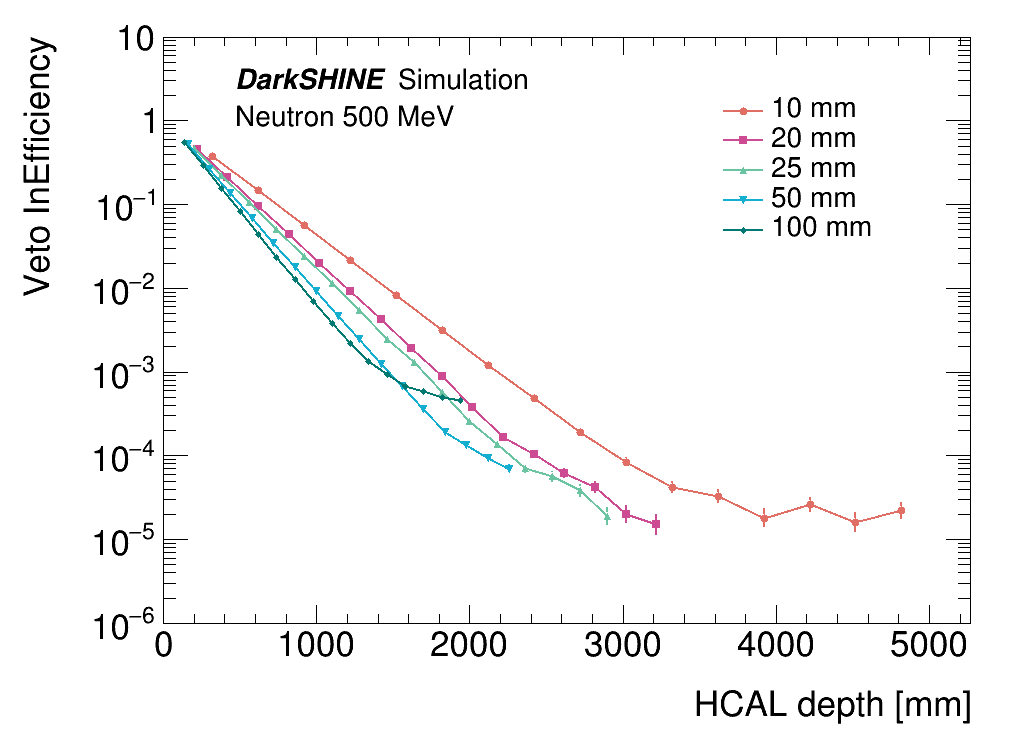}
\includegraphics
  [width=0.45\hsize]
  {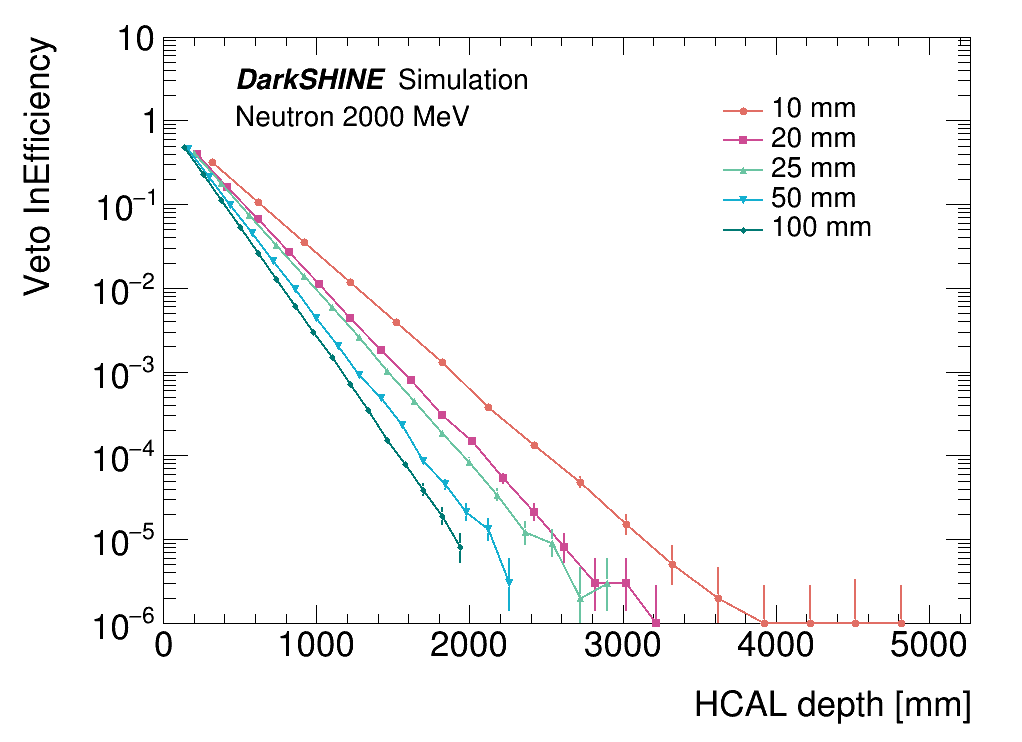}
\caption{Veto inefficiency as a function of detector depth for different absorber thickness. Here, 100, 500, and 2000 MeV neutrons are generated to hit towards hadronic calorimeter at its center. The veto efficiency of low-energy neutrons can be enhanced in thinner absorber thickness, while thicker absorber thickness enables the vetoing of high-energy neutrons within a smaller depth range.}
\label{fig:ineff_thickness}
\end{figure*}

This study simulated various absorber thicknesses within each detecting unit, ranging as 10–100 mm per layer, for different cases. The veto inefficiency as a function of the detector depth is illustrated in figure~\ref{fig:ineff_thickness}, where the three plots represent the tests conducted with neutron injections at energies of 100, 500, and 2000~MeV. Each curve in these plots has the same number of test points representing the total absorber thickness from 100 to 1600mm, incremented by 100mm per step. The overall HCAL depth was used as the x-axis instead of the absorber thickness because it was easier to correspond the x-axis numbers to the overall HCAL dimensions. \\

\begin{figure*}[htb]
\centering
\includegraphics
    [width=0.45\textwidth]{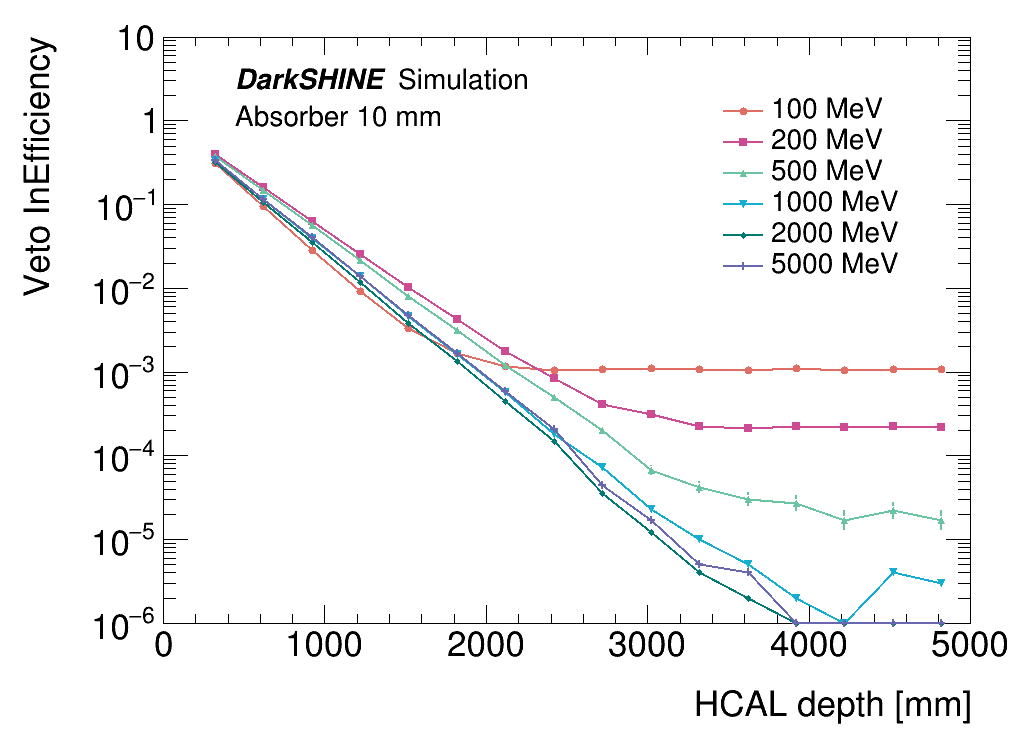}
\includegraphics
    [width=0.45\textwidth]{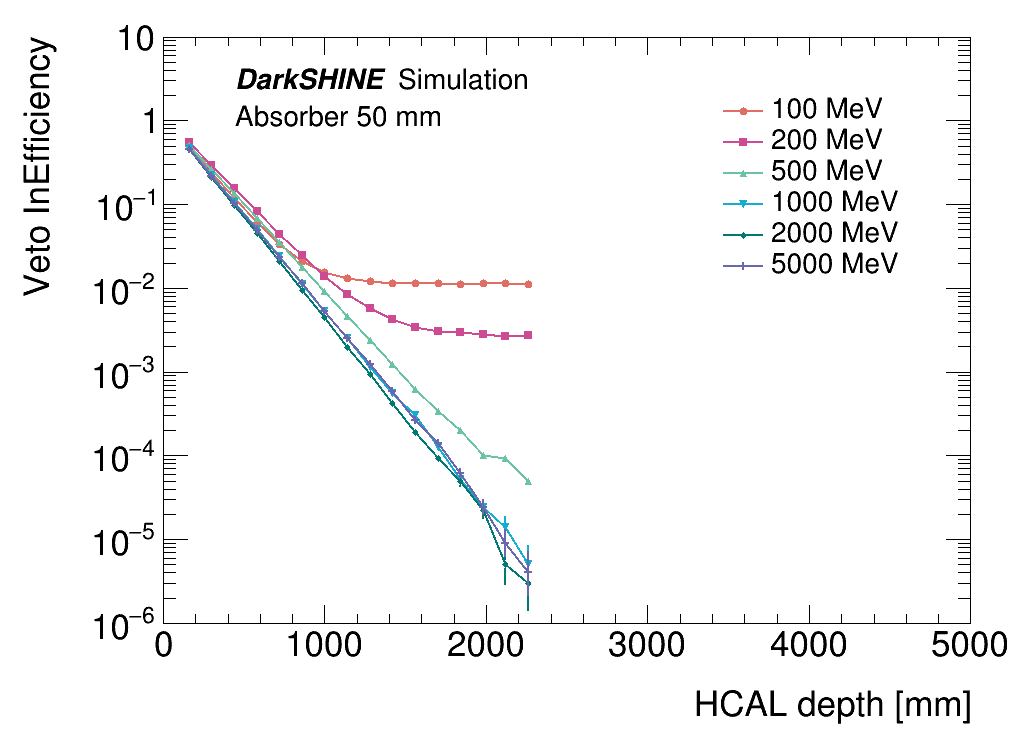}
\caption{Veto inefficiency as a function of detector depth for different beam energy. Here, 10 and 50~mm absorber thickness results are shown. All the curves have same number of points, and the corresponding points in two plots represent same total absorber thickness but different HCAL depth.}
\label{fig:ineff_energy}
\end{figure*}

The performance of a 10mm absorber is optimal for 100MeV neutrons, achieving a platform with approximately 70 layers. Both thick and thin absorbers can attain an inefficiency of $<$10$^{-5}$ for high-energy neutrons when provided with sufficient layers; however, thicker absorber requires fewer layers when the depth remains constant. To understand these scenarios more clearly, the veto inefficiencies as a function of the detector depth for different beam energies with two choices of thicknesses, 10 and 50~mm, are shown in Figure~\ref{fig:ineff_energy}. With an increased thickness of the absorber, it is feasible to reduce the number of layers and achieve an inefficiency of $<$10$^{-5}$, as demonstrated in this study. \\

It can be concluded that the new design for the DarkSHINE HCAL involves combining a thinner absorber in the front half of HCAL to collect deposit energy from low-energy neutrons and a thicker absorber in the remaining part to minimize the total material used; this approach employs 70 layers of a 10 mm absorber and ~18 layers of a 50 mm absorber, which together achieve a total thickness of approximately 10 $\lambda$, thus satisfying the physical requirements. \\

\subsection{Scintillator layer design}
\label{sec:xAbsy}
The scintillator layer comprises multiple scintillator strips, which are extensively employed in calorimeters~\cite{Luo:2023inu} along with a photomultiplier device, such as SiPM~\cite{Hukun,Sipm}. However, owing to packaging and mechanical constraints, there exist inevitable gaps resembling a fence. By rotating the second scintillator layer by 90°, the scintillator strips became orthogonal to those in the first layer, effectively complementing each other's gaps. A schematic of its basic structure is shown in Figure~\ref{fig:xy_crossing}. However, it remains uncertain whether an intermediate structure between every two absorbers is necessary or if a single layer followed by an absorber would suffice. \\

\begin{figure*}[]
\centering
\includegraphics
  [width=0.45\hsize]
  {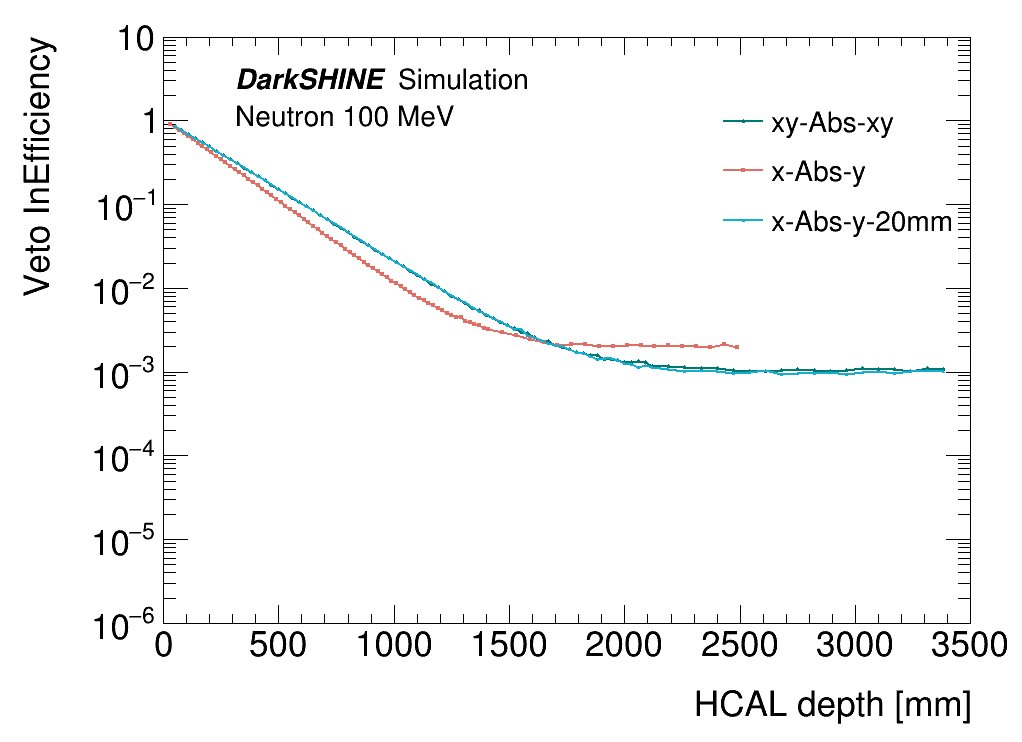}
\includegraphics
  [width=0.45\hsize]
  {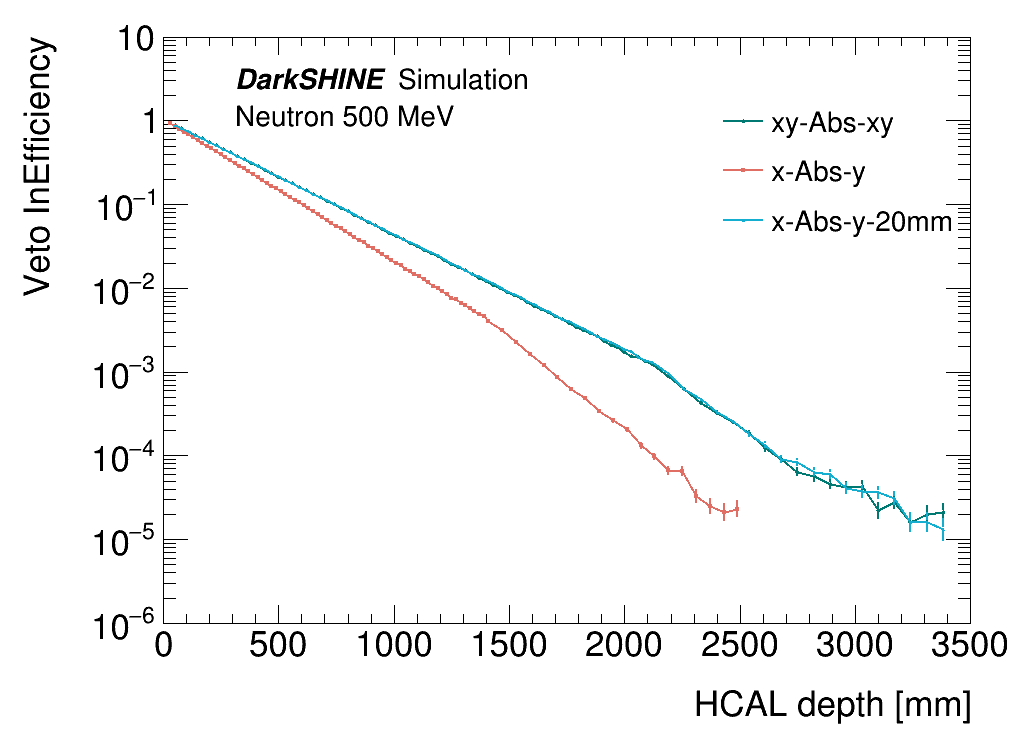} 
\includegraphics
  [width=0.45\hsize]
  {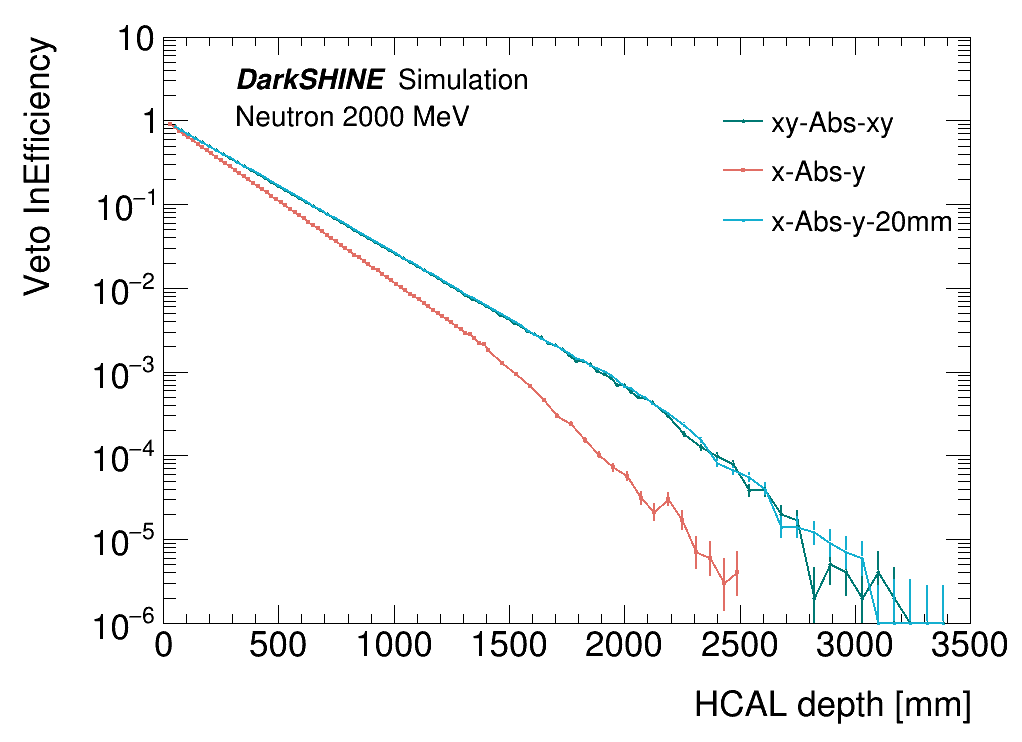}
\includegraphics
  [width=0.45\hsize]
  {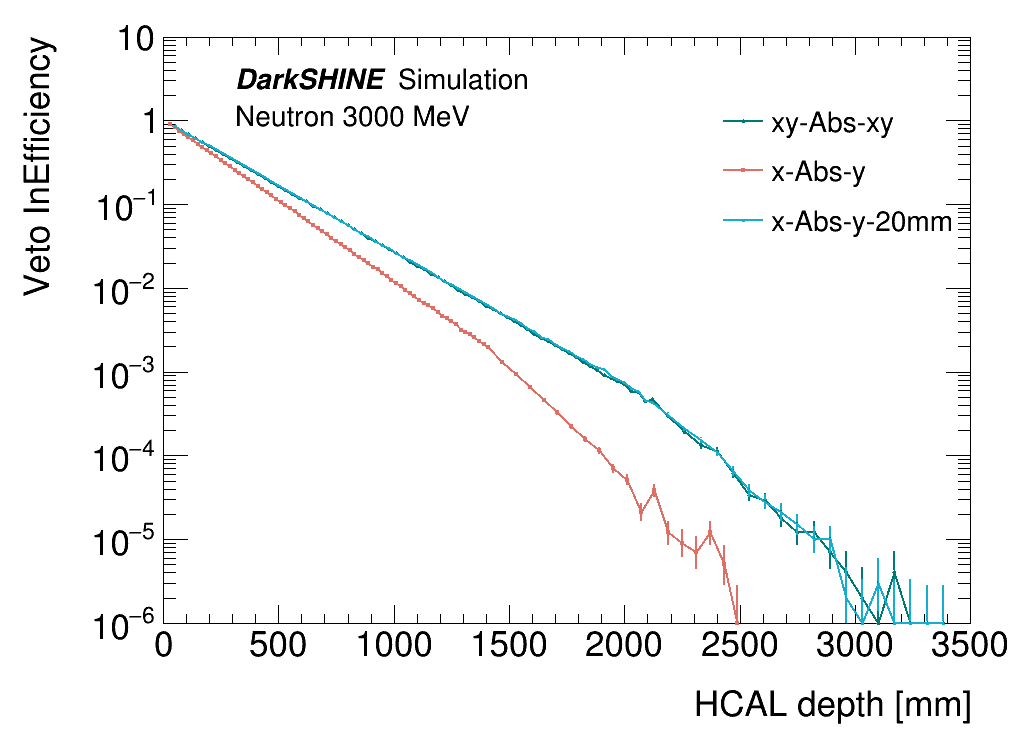} 
\caption{Veto inefficiency as a function of detector depth for different absorber thicknesses and scintillator strategy. Here, 100, 500, 2000, and 3000~MeV neutrons are generated to hit towards hadronic calorimeter at its center. Further, 10~mm and 50~mm absorbers are used in the first 70 layers and the remaining part, respectively. The 20~mm-scintillator in x-Abs-y design exhibits approximately the same performance as the 10~mm-scintillator in xy-Abs-xy design. The 10~mm-scintillator in the x-Abs-y design performs slightly worse in low-energy scheme but reduces the amount of scintillator.}
\label{fig:ineff_aAbsy}
\end{figure*}

The present study was conducted based on the results of the absorber thickness optimization presented in Section~\ref{sec:Absorber}, wherein the initial 70 layers utilize a 10~mm absorber, followed by 50~mm layers. A comparison of veto inefficiencies is visualized in Figure~\ref{fig:ineff_aAbsy}. The default configuration involves the insertion of two 10~mm layers between each pair of absorbers, designated as `xy-Abs-xy,' where `xy' denotes the pair of scintillator layers and `Abs' represents absorber. An alternative configuration is labeled `x-Abs-y,' which signifies the insertion of an absorber layer between every two 10~mm scintillator layers with orthogonal strip direction. Additional comparisons were also performed. The thickness of the scintillators was set to 20mm. \\

All the configurations exhibit satisfactory performance for high-energy neutrons. For 100 MeV neutrons, a 10~mm thick x-Abs-y configuration exhibits a slightly inferior performance; however, considering that the overall veto efficiency is the product of the inefficiencies of all particles in one event, it remains acceptable. The 20~mm thick x-Abs-y configuration demonstrates virtually identical performance to the xy-Abs-xy configuration, which is consistent with expectations, but does not reduce scintillator consumption. The x-Abs-y design with a 10~mm thick scintillator utilizes half the materials compared with the 20~mm configuration. \\

\subsection{Optimization performance}
\label{sec:performance}

As stated in Section~\ref{sec:background_process}, a previous study utilized the neutron veto inefficiency as the primary research indicator; however, it is imperative to evaluate the performance of all relevant particles entering HCAL. The veto inefficiency of various hadronic particles traversing the optimized DarkSHINE HCAL is assessed by simulating different incident particles with distinct energies. The number of events was increased to 10$^7$ and the test was conducted on neutrons, k$^0$ (both short- and long-lived), $\pi^0$, and protons. The results are presented in Table~\ref{tab:hadronic_particle_ineff}. \\

The HCAL demonstrated a nuanced performance gradient in its veto capabilities, particularly when examining its interaction with neutrons of disparate energies. Initial observations revealed that neutrons in the lower-energy spectrum, approximately 100 MeV, were predominantly absorbed within the first few calorimeter's absorber layers. This absorption effectively precluded the generation or detection of secondary particles by the HCAL's sensitive layers. Such phenomena underscore the challenges inherent in detecting lower-energy neutrons, owing to their minimal interaction with detector materials. \\

As we progress to a higher energy threshold, specifically approximately 500 MeV, there is a notable decrease in the veto inefficiency. This decrease is indicative of the HCAL's enhanced capability to effectively identify and veto neutron events, aligning with the theoretical predictions and design objectives. The underlying mechanism facilitating this improved performance is likely related to the increased production of secondary particles and their subsequent energy deposition within the sensitive layers of the HCAL, thereby enabling more efficient detection and vetoing of incident neutrons. \\

Advancing further into the energy scale, particularly in realms exceeding 1000 MeV, the HCAL shows a consistently low veto inefficiency across a broad array of hadronic particles, including but not limited to neutrons, kaons, and pions. The inefficiencies observed for these particles generally reside within the order of magnitude of -5, illuminating the HCAL's superior discrimination capabilities against higher-energy hadronic particles. This enhanced performance can be attributed to the calorimeter's design optimizations, which were specifically tailored to maximize its sensitivity and specificity in the higher-energy domain. \\

In addition to the overall trend of decreasing veto inefficiency with increasing energy, it was observed that the veto inefficiency of kaon was higher at 2~GeV. This can be attributed to the performance characteristics of long-lived K$^0$-particles. As the energy increases, showers tend to occur deeper within the detector, and secondary particles such as neutrons, which are more challenging to detect, also possess higher energies. Consequently the effectiveness of the veto is slightly reduced. \\

Moreover, the differential behavior exhibited by various particle types within the HCAL elucidates the complex interactions between the particles and detector materials. The variability in veto inefficiency among different hadrons reflects their distinct interaction mechanisms within the calorimeter, highlighting the sophisticated nature of the HCAL's operational parameters. These findings validate the efficacy of the optimized HCAL design in satisfying the DarkSHINE experiment's stringent requirements and provide valuable insights into the fundamental aspects of particle detection and background event mitigation in high-energy physics research. \\

\begin{table}[!htb]
\caption{Veto inefficiency of HCAL targeting different incident hadronic particles with different energies. Events with multiple hadronic particles are more easily vetoed by HCAL detector under assumptions that veto power of different particles at one event is independent.}
\label{tab:hadronic_particle_ineff}
\centering
\renewcommand{\arraystretch}{1.3}
\begin{tabular}{|c|c|c|}
\hline
Veto InEff \(\times\)E-06 & n & \(k^0\) \\ \hline
100[MeV] & 1170\(^{+10.9}_{-10.8}\) & 31600\(^{+55.5}_{-55.4}\) \\ \hline
500[MeV] & 18.4\(^{+1.46}_{-1.36}\) & 5.40\(^{+0.839}_{-0.733}\) \\ \hline
1000[MeV] & 3.70\(^{+0.714}_{-0.606}\) & 3.70\(^{+0.714}_{-0.606}\) \\ \hline
2000[MeV] & 2.70\(^{+0.626}_{-0.516}\) & 11.5\(^{+1.19}_{-1.08}\) \\ \hline
\end{tabular}

\vspace{0.2cm}

\begin{tabular}{|c|c|c|}
\hline
 \(\pi^0\) & p & \(\mu\) \\ \hline
7.30\(^{+0.958}_{-0.852}\) & 30700\(^{+61.5}_{-61.3}\) & 409\(^{+6.49}_{-6.39}\) \\ \hline
0.1\(^{+0.184}_{-0}\) & 8.04\(^{+1.34}_{-1.16}\) & 15.0\(^{+1.33}_{-1.22}\) \\ \hline
0.1\(^{+0.184}_{-0}\) & 0.1\(^{+0.958}_{-0}\) & 2.00\(^{+0.555}_{-0.443}\) \\ \hline
0.1\(^{+0.188}_{-0}\) & 0.1\(^{+2.78}_{-0}\) & 0.1\(^{+0.184}_{-0}\) \\ \hline
\end{tabular}

\end{table}

\section{Side HCAL design}
\label{sec:sideHcal}

The optimization and validation of the target particle rejection power have been demonstrated, and this power is equivalent to the rejection power of background events where particles are incident on the HCAL. However, not all these events occur within this context. Particles can exit the ECAL via alternative paths and fail to enter the HCAL. In the rare processes discussed in Section ~\ref{sec:background_process},  events may involve secondary particles with substantial azimuthal angles relative to the beamline. These processes, such as photon-nuclear and electron-nuclear interactions, can occur at various locations within the detector system. Secondary particles emitted by these interactions may be overlooked if the detector cannot cover a specific solid angle. Based on a study of 1$\times$10$^{8}$ electron-on-target events, it was found that only 65\% of the neutrons in these events could reach the first layer of HCAL within a 150$\times$150~cm$^2$ area. The detailed numbers are presented in Table~\ref{tab:n_region_ratio}, considering both 50$\times$50~cm$^2$ (directly entering HCAL from the end of ECAL) and 150$\times$150~cm$^2$. The results with and without a $\mathrm{E_{ECAL}} < 2.5$~GeV cut are provided for comparison. \\

\begin{table}[!htb]
\caption{Ratio of neutrons inside and outside the 50$\times$50~cm$^2$ and 150$\times$150~cm$^2$ regions of the HCAL.}
\label{tab:n_region_ratio}
\centering
\begin{tabular}{|c|c|c|}
\hline
\diagbox{Cut}{Size} & 50$\times$50~cm$^2$ & 150$\times$150~cm$^2$ \\ \hline
No cut & 0.08 & 0.43 \\ \hline
$\mathrm{E_{ECAL}} < 2.5$~GeV & 0.23 & 0.65 \\ \hline

\end{tabular}
\end{table}

Therefore, an additional segmented HCAL, referred to as a side HCAL, was incorporated. The side HCAL comprises four cuboid parts that encircled and enveloped the ECAL end-to-end with its sensitive surface perpendicular to the ECAL. The width of the proposed structure is envisioned to align with the ECAL's depth in the z direction, whereas its length (in either the x or y direction) corresponds to the sum of ECAL's width and half the difference between HCAL's transverse size and ECAL's lateral length, representing the distance from one side of ECAL to the furthermost side of HCAL. \\

The side HCAL is also composed of multiple iron absorber and plastic scintillator layers, each with an area of 45~cm$\times$105~cm, which correlates to the ECAL dimensions~\cite{Chen:2022liu}. Owing to the non-square layer shape, the 'x-y' design would result in differently sized scintillator strips and is thus not employed. Furthermore, the uneven absorber thickness is superfluous because the depth is significantly smaller than that of the main body. Each sensitivity and absorber layer was 10mm thick, collectively comprising 50 layers. \\

The performances of the two designs, with and without the side HCAL, were evaluated separately. Simulated events of four types of rare processes were employed to investigate veto inefficiency. At this stage, the combination of optimized designs discussed in our study was implemented with no additional subdetector cuts employed. These numbers should not be considered a comprehensive measure of the DarkSHINE experiment's veto inefficiency in excluding rare process events, but rather serve as a means to explore additional related events for comparative analysis within this specific context. The results are presented in Table~\ref{tab:ineff_structure_rare_process}, indicating that the inclusion of a side HCAL significantly enhances the overall performance. \\

\begin{table}[!htb]
\caption{Veto inefficiency by simulating 8-GeV electron-on-target events into different structures and with different rare process biased. Based on the numbers, designs with side HCAL exhibit better veto power than designs without side HCAL.}
\label{tab:ineff_structure_rare_process}
\centering
\begin{tabular}{|c|c|c|c|c|}
\hline
\diagbox{Structure}{Process}& EN-target & EN-ECAL  & PN-target & PN-ECAL   \\ \hline
w/o Side HCAL  & 2.68E-02 & 3.94E-02 & 9.29E-02 & 1.24E-01 \\ \hline
w/ Side HCAL   & 1.04E-03 & 1.09E-02 & 1.94E-03 & 3.58E-02 \\ \hline
\end{tabular}

\end{table}

\section{Conclusion}
\label{sec:conclusion}

The design and optimization of the hadronic calorimeter for the DarkSHINE experiment are conducted in detail. Factors such as the thickness of the absorber, placement of the scintillator, and requirements of the side HCAL, which could influence the veto performance, have been thoroughly investigated. Several crucial parameters are identified for optimization, considering constraints such as limited weighting and budget conditions, which also satisfy the physical requirements. \\

In the optimized design, the HCAL comprises iron absorber and scintillator sensitive layers. The transverse dimensions are 1.5 m $\times$ 1.5 m, with approximately 10 $\lambda$ thick iron absorbing layers along the beam direction. The first 70 layers have a thickness of 1cm, whereas the last 18 layers are 5cm thick. Following each absorber, there is a 1~cm thick plastic scintillator layer composed of a strip measuring 5cm in width. The scintillator strips in the two adjacent layers before and after the absorber layer were oriented perpendicular to each other. Additionally, a scintillator layer is positioned between the ECAL and the first iron layer. In addition to the main HCAL, a side HCAL surrounding the ECAL is incorporated. \\

The novel design of HCAL offers enhanced veto power for various rare processes, including electron-nuclear and photon-nuclear reactions, particularly those with neutral hadrons and muons in the final decay states. This optimized HCAL design is not only crucial for the DarkSHINE experiment but also demonstrates the importance of advanced hadronic calorimetry techniques in particle physics experiments of this kind. \\
\\
\\

\end{document}